\documentclass[fleqn,12pt]{wlscirep}

\usepackage{cite}

\newcommand{\be}{\begin{equation}}
\newcommand{\ee}{\end{equation}}
\newcommand{\bea}{\begin{eqnarray}}
\newcommand{\eea}{\end{eqnarray}}
\def \ket #1{\left| #1 \right\rangle}

\def \scalarprod #1#2{\left \langle #1 \right. \left| #2 \right\rangle}
\def \x{\mathbf{x}}
\def \r{\mathbf{r}}

\begin{document}

\title{ Selective final state spectroscopy and multifractality in two-component ultracold Bose-Einstein condensates: a numerical study}

\author[1,*]{Mikl\'os Antal Werner}
\author[2]{Eugene Demler}
\author[3]{ Alain Aspect}
\author[1]{Gergely Zar\'and}
\affil[1]{Exotic Quantum Phases ``Momentum'' Research Group,
Department of Theoretical Physics, Budapest University of Technology and Economics, 1111 Budapest, Budafoki út 8, Hungary}
\affil[2]{Department of Physics, Harvard University, Cambridge, Massachusetts 02138, USA}
\affil[3]{Laboratoire Charles Fabry
Institut d’Optique Graduate School – CNRS – Université Paris Sud,
2 avenue Augustin Fresnel, 91127 Palaiseau, France}
\affil[*]{werner@phy.bme.hu}


\begin{abstract}
{We propose to use the method introduced by Volchkov et al., based on state dependent disordered   ultracold bosons, to address the critical state at the mobility edge 
of  the Anderson localization transition, and to observe its  intriguing multifractal structure.   
 An optimally designed external radio frequency  pulse  can be applied  to generate transitions to eigenstates in a  narrow energy window 
 close to  the mobility edge, where  critical scaling and multifractality emerge. 
 Two-photon laser scanning microscopy
will be used to address individual  localized states even close to the transition. 
 The projected image of the cloud is shown to 
 inherit multifractality and to display universal density correlations.  
  Time of flight images  of the excited states are  predicted to show interference fringes in 
 the localized phase, while they allow one to map  equal energy surfaces  deep in the metallic phase.}
\end{abstract}

\flushbottom
\maketitle
%
%
\thispagestyle{empty}

\section*{Introduction}
Anderson localization is one of the most fundamental quantum interference phenomena in disordered 
quantum systems. As first pointed out in the seminal work of Anderson\cite{Anderson}, 
all eigenstates of a particle on 
a disordered lattice  localize in space in sufficiently strong disorder. Though this statement is independent of dimensionality, the existence of delocalized 
states and the nature of eigenstates at various energies still depend on the spatial dimension of the system considered, and may also be affected by the specific 
structure of the disorder\cite{Sanchez-Palencia:2007rj,
Lugan:2009ma,gurevich2009lyapunov}.
While in $d=1$ dimension all eigenstates are always localized for uncorrelated diagonal disorder,
in  dimensions $d>2$ a critical disorder strength exists\cite{Review1, Review2}. Below this critical disorder,
  mobility edges $E_{\rm mob}$ separating localized and delocalized states emerge.
  At the critical energies, $E=E_{\rm mob}$, a peculiar quantum phase transition takes place.
Approaching  $E_{\rm mob}$ through localized states, the typical spatial extension $\xi$ of the wave functions
is found to diverge as a power law, $\xi(E)\sim |E-E_{\rm mob}|^{-\nu}$, with $\nu$ a universal  exponent that depends only on dimensionality 
and the symmetry of the underlying Hamiltonian\cite{Abrahams1979}. In case of $d=3$ spatial dimensions and in the absence of external magnetic fields and spin-orbit 
interaction, studied here,  the exponent  $\nu=\nu_{\rm orth}=1.58\dots$ has been determined with great accuracy by transfer matrix methods, with the 
label 'orth'   referring to orthogonal universality class\cite{Ohtsuki}.  
 At the critical energies, $E=E_{\rm mob}$,   eigenstates of universal properties emerge;
the absence of length scale, $\xi\to \infty$  implies   a self-similar character and  entails  a 
multifractal structure\cite{MF_book, MF_review} of the wave function\cite{SoukoulisEconomou83, SchreiberGrussbach, MirlinEvers2000,Kravtsov1}. 
In particular, at criticality, the probability distribution 
associated with the 
critical wave function $\psi_\r$ is predicted to scale  with system size as
\be
P \left( |\psi_\r|^2 \sim L^{-\alpha} \right) \propto L^{{f}(\alpha)-d} \;,
\label{eq:multifractal}
\ee
with $f(\alpha)$ the universal multifractal spectrum of the critical wave function and $L$ the system's linear extension\cite{RomerPRL2009, MirlinReview}.
This multifractal structure has a deep origin: it is interpreted as  a signature of the presence of infinitely many relevant operators
at the Anderson transition\cite{MirlinReview}. Remarkably, multifractality is also suggested to emerge in the context of many-body localization: 
recent numerical studies  of the Anderson problem on a random regular graph (RRG) reported the existence of the non-ergodic phase 
with  eigenfunctions exhibiting  multifractal structures  with disorder dependent  fractal dimensions\cite{Biroli2012,DeLuca2014,Altshuler2016}. 

Though Anderson's  localization transition has been 
experimentally studied
in a range of systems including 
doped semiconductors\cite{Choi}, granular optical media\cite{Wiersma,van2012surface,storzer2006observation,sperling2016can}, 
disordered microwave waveguides\cite{MW_localization},
 elastic networks\cite{vanTiggelen_ultrasound}, photonic lattices\cite{Segev,Lahini},
and cold atomic systems\cite{Aspect_1D, Inguscio_1D,Aspect,DeMarco_Science2011,Semeghini2015}, in spite of the considerable effort\cite{vanTiggelen_MF,Yazdani,Morgenstern},
 a convincing measurement of the critical state's  predicted multifractal spectrum remains 
 missing. 

In conventional solid state systems such as doped semiconductors, e.g., Scanning Tunneling Microscopy (STM)
 provides a way to gain information on the detailed structure of the critical wave function\cite{Morgenstern,Yazdani}. 
 Indeed, in doped ferromagnetic semiconductors,  emergent short distance power law 
signatures  have been demonstrated close to criticality at the Fermi energy\cite{Yazdani}, but the predicted  anomalous exponent has not been observed and  the computed multifractal spectra remained far from the 
theoretically predicted universal form. Rather, measured amplitude distributions exhibited close to lognormal distributions, characteristic  of disordered metals. Moreover, in these systems, strong electron-electron interactions have a large impact 
on the metal-insulator transition, and Anderson's non-interacting theory looses its 
 validity\cite{Finkelstein_etc,Mueller}. 
 STM signatures of  multifractal properties  have   also been reported  on iron-doped InAs surfaces in the context of quantum-Hall transition\cite{Morgenstern}, 
but the energy resolution was insufficient to observe the mulifractal spectrum of the 
critical quantum-Hall state\cite{Huckestein_review}.  Indications of multicriticality have also been reported in ultrasound experiments\cite{vanTiggelen_MF}, 
but to our knowledge the critical mulifractal spectrum has never been observed in these experiments, either.

Experiments with ultracold atoms   provide powerful and versatile means to realize and study 
the Anderson transition. In these experiments  laser speckles\cite{Aspect,DeMarco_Science2011}, 
quasiperiodic optical lattices\cite{Inguscio_1D},  or holographic methods\cite{GreinerHolographic, HolographicPot, DeMarco} 
can be used to produce  disorder, and the 
interaction of the atoms can  be controlled by changing the depth of an applied  optical lattice or 
the strength of the confinement potential or using Feschbach resonances. 
Localization transition has   been observed in
one- and three-dimensional systems\cite{Aspect_1D, Inguscio_1D,Aspect,DeMarco_Science2011,Semeghini2015}, and weak localization and coherent back scattering  
in two dimensions  has also been investigated experimentally\cite{Aspect_backscattering,muller2015suppression}. 
None of these experiments  could, however,  detect the critical state so far. In fact, detecting the critical state turns out to be a very delicate task: 
importantly, the multifractal behavior is  the property of a \emph{single } eigenstate and, as we shall see,  admixture of 
just a few  eigenstates by interactions or inelastic processes even in a narrow energy window makes it hard to observe.

\begin{figure}
 \centering
 \includegraphics[width=0.8\textwidth]{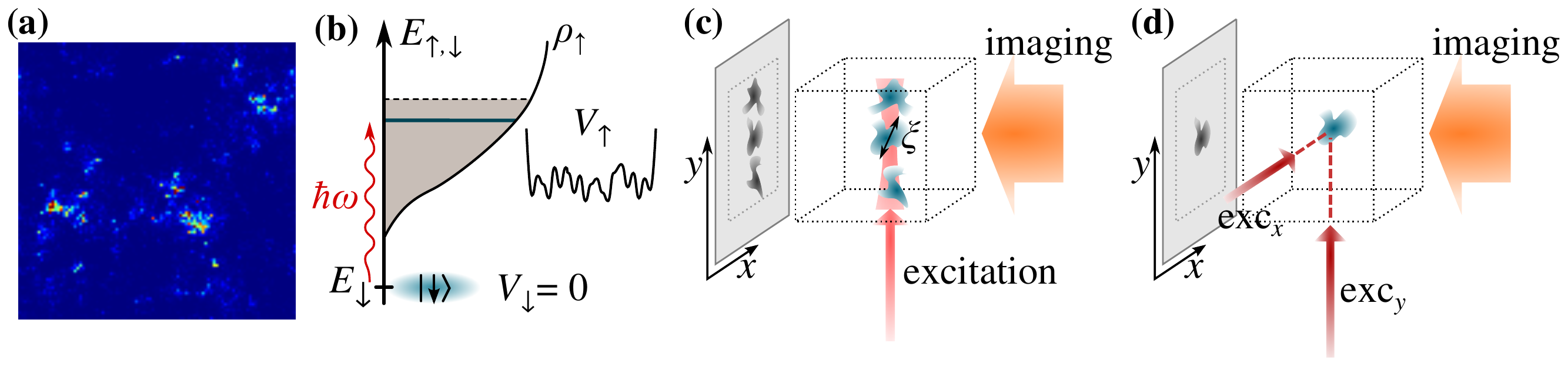}
 \caption{(a) Projected image of a critical multifractal eigenstate of size $(L/a)^3= 40^3$, exhibiting clustering. (b-c) Schematics of the proposed experiment: (b) The condensate is prepared in the hyperfine state $\ket{\downarrow}$ where no disorder is present. The external field excites
 the atoms to the hyperfine state $\ket{\uparrow}$, interacting with the disorder potential $V_\uparrow$. The energy $E_\uparrow$  
 of the final state is controlled by the frequency $\omega$. 
 (c) Excited atoms in the $\ket{\uparrow}$ state are imaged by a horizontal laser beam. The overlap between different eigenstates can be reduced  by using a 
 narrow excitation beams of waist $w_0\lesssim \xi$, where $\xi$ is the typical size of a localized state. (d) Even a single localized state can be excited and imaged by crossing two laser beams. }\label{exp_setup}
\end{figure}

Here we propose  to image the critical state and to detect its multifractal properties in  peculiar cold 
atomic setups, sketched in Fig.~\ref{exp_setup}. Following the method proposed and demonstrated by Volchkov et al\cite{2017arXiv170707577V}, we suggest  using  bosons with two hyperfine components, 
$\sigma = \uparrow$ and $\downarrow$, with only the component $\uparrow$ being subject to a random potential 
$V_\uparrow$. Placing  all bosons initially in the weakly interacting (and homogeneous) state $\downarrow$ 
and exciting from 
this spin $\downarrow$ condensate spin $\uparrow$ states with the external frequency $\omega$, 
the energy $ E_\uparrow =  E_\downarrow + \hbar \omega$  of the final state can be  selected   and tuned 
through the mobility edge of the spin $\uparrow$ component either by changing $\omega$ or by varying the disorder amplitude. 
As we demonstrate through detailed large scale simulations,  it is possible to single out only a few final states in a very narrow energy window by (1) 
carefully optimizing the shape of the  excitation pulse, (2) switching off interactions in the final state,   and (3) by using carefully designed excitation geometries. 

In particular, we propose to use  two different arrangements. In the first geometry, we propose to use  
a thin vertical  laser beam with  two frequencies
to produce two-photon Raman  transitions to  localized states in a narrow spatial range and   at a well-defined energy
  such that that the excited wave functions do not overlap.  Projection imaging along the horizontal direction
  (see Fig.~\ref{exp_setup}c) yields images of non-overlapping localized wave functions close to the mobility edge $E_{\rm mob}$.
Alternatively, instead of   one thin vertical  beam with two frequencies, one 
could use \emph{crossed} laser  beams to address \emph{individual  localized} states via two-photon Raman transitions 
in just the small crossing region (see Fig.~\ref{exp_setup}d).  This method, developed in biological microscopy and known as `Two Photon Laser Scanning Microscopy' (TPLSM)\cite{xu1996multiphoton},
is similar in spirit to the one proposed by Kollath \textit{et al.}\cite{Kollath2007}, but here no probe atoms are needed. Rather, transitions are generated 
from a Bose condensate of atoms, and the local character of the transitions is ensured through  crossing the two laser beams.  

As we show below through detailed simulations,  multifractal properties 
of the critical state are  \emph{inherited} by the projected pictures of  
localized states with sufficiently large  localization 
lengths. The projection of the squared amplitude of  wave function, i.e., the density exhibits a non-trivial multifractal spectrum which we determine 
through large scale simulations, and the 
 projected  density-density correlation  functions are predicted and demonstrated to display  universal scaling collapse, thereby 
 providing evidence for multifractal behavior.

\section*{Results}

In the absence of radiation, we can describe the two-component Bose mixture in the optical lattice by the Hamiltonian, 
\be\label{eq:Hamiltonian}
{H} = - J\sideset{}{'} \sum_{\r,\r',\sigma} 
a^\dag_{\r\sigma} a_{\r'\sigma} + 
	  \sum_{\r,\sigma} \varepsilon_{\r \sigma}  a^\dag_{\r\sigma} a_{\r\sigma}
 +  \sum_{\r,\sigma,\sigma'} \frac{1}{2} U_{\sigma \sigma'} a^{\dag}_{\r\sigma} a^{\dag}_{\r\sigma'} a_{\r\mathbf{\sigma}'} a_{\r\sigma},
\ee
with the prime indicating summation  over neighboring sites only. The onsite energies $ \varepsilon_{\r \sigma}$ 
incorporate the chemical potential, and $ \varepsilon_{\r \uparrow}$   also contains the random component $V_\uparrow(\r)$, 
responsible for Anderson localization of the $\uparrow$ bosons.
In practice, the random potential $V_\uparrow(\r)$
is realized  by  fine-grained speckle 
potentials\cite{DeMarco_PRL2009} or holographic  disorder\cite{DeMarco}. 
Here, for simplicity, we shall replace it by a uniform and independent  distribution of $  \varepsilon_{\r \uparrow}\in[-W /2,W /2]$ 
on each lattice site.

A possibility to generate hyperfine $\uparrow\;\Rightarrow\;\downarrow$ transitions in a sufficiently narrow spatial region 
is to  use two-photon processes\cite{xu1996multiphoton}. Here we consider a stimulated Raman process\cite{2017arXiv170707577V}, which 
generates an effective RF field with a frequency  corresponding   to the beating frequency   of the two lasers, 
and can be described in the rotating frame  approximation by the 
coupling term
\be
H_\Omega = \sum_{\r} \hbar \Omega_{\r}(t)  \;
( a^\dag_{\r \uparrow} a_{\r \downarrow} + h.c.),
\ee
with $\Omega_\r(t)$ the Rabi frequency, and the RF frequency appearing simply as an energy shift 
$\varepsilon_{\uparrow,\downarrow}\to \varepsilon_{\uparrow,\downarrow}\pm \hbar\omega/2 $. For $\Omega_\r(t)$ we assume a 
Gaussian profile, 
$\Omega_\r\sim \Omega_0(t) e^{-2(x^2+y^2)/{w_0}^2}$ with a narrow waist ${w_0}$ in the range of a few lattice constants, ${w_0}\sim 2\;{\rm \mu m}$
The shape of the pulse $\Omega_0(t)$ must be determined carefully to generate transitions between the two hyperfine components in 
an energy window as narrow as possible. 
Since the coupling happens between a discrete state and a quasi-continuum, the width of the final energy window 
follows from Fermi's Golden rule\cite{IntroToQuantOpt}, and can be adjusted by the duration and shape of the pulse $\Omega(t)$.
However, interactions in the final state turn out to be particularly destructive: they generate unwanted 
transitions and broaden the spectrum of the final state. 
Therefore, in the following, we shall assume that interactions in the final state have been switched off 
 by applying an appropriate external magnetic field, $U_{\uparrow\uparrow=0}$.  
We assume further that the two hyperfine states are decoupled  for $t<0$, 
 $\Omega_0(t<0)=0$, and the bosons  form a condensate  in the  state $\downarrow$, 
 described by a collective wave function $a_{\r,\sigma}\to\psi_{\r\sigma}$ at time $t=0$. We describe  the time 
 dependence of this  collective wave function in terms of the Gross-Pitaevskii equation (see Methods).

\subsection*{Numerical simulations.}
We determined the time dependence of the collective wave function by solving  
Eq.~\eqref{eq:GPequation} numerically. To maintain numerical accuracy, we suppressed
 fast phase oscillations of the wave function by means of a self-consistently determined 
global gauge transformation (see Eq.~\eqref{eq:GPgauge} in Methods). This allowed us to perform
Gross-Pitaevskii simulations on systems  as large as  $40\times 40\times 40$, where length is measured in units of the lattice constant $a$. 
Since the system size  is not larger than 
the Rayleigh lenght $\pi w_0^2 / \lambda$ of the laser beam, we neglected the spatial spreading of the waist, and assumed a simple cylindrical beam. The  RF field $\Omega_{\r}(t)$ was 
 turned on smoothly, kept constant and then turned off with an optimized pulse  of length 
$T = 400 \;{\hbar}/{J}$ and turn on  time $\tau = 60 \;{\hbar}/{J}$. This resulted in an energy window $\Delta \epsilon$ of the final states 
encompassing a limited number of states (see the discussion below).

\begin{figure}
\centering
 \includegraphics[width=0.7\textwidth]{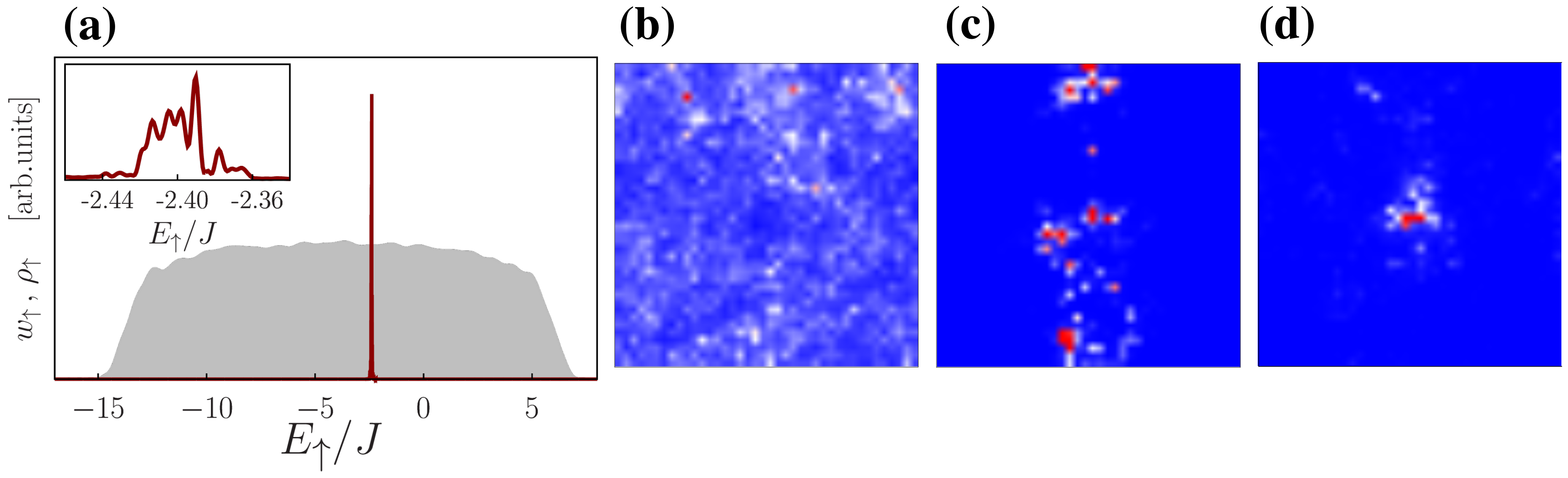}
 \caption{
 \label{fig:energy_resolution}
 (a) Spectral weight $w_{\uparrow}(E_\uparrow)$ of the bosons as a function of the final state's energy $E_\uparrow$ after the excitation pulse for a lattice of $(L/a)^3= 40^3$ sites,  $W=17J$ and 
 excitation beam waist $w_0=4a$. We simulated $N=3200$ atoms with 
 typical interaction strengths  $U_{\uparrow\downarrow} = U_{\downarrow\downarrow} = 2J$, excitation 
 frequency $\omega  \approx 8 J/\hbar$.  The grey area indicates the density of states. 
 Inset: same spectral peak on a smaller scale. (b - c) Projected images of a metallic ($W=10J$) and a 
 localized ($W = 28J$) final state. The excitation beam has vertical direction in both cases. 
(d) Projected  image of a single localized state, excited  by two crossed laser beams.
All figures show  images of size $40a\times 40a$. }
\end{figure}
In small disordered systems, disorder averaging is necessary to  eliminate sample to sample fluctuations.  
Averaging over a huge ensemble of disorder realizations is  experimentally demanding.
Disorder averaging can, however, also be replaced  in part by  averaging  over the RF frequency  $\omega$  
in a small window, since  the localization length $\xi(E)$ displays  only a weak energy dependence near the 
band center\cite{Review1}.

The spectral and spatial structure of the state reached after the excitation pulse is shown in 
Fig.~\ref{fig:energy_resolution} for some typical parameters. As shown in panel (a), states are excited in a very 
narrow energy window. 
To determine the spectral weights 
$w_\uparrow(E)$ and the density of states, displayed in panel (a), we employed the so-called 
Chebyshev polynomial expansion (see Methods),  which made possible to obtain high 
accuracy results and thus circumvent  the impossible task of performing a complete 
diagonalization of the final Hamiltonian.  Although 
states in a very narrow window of width $\Delta \epsilon \sim 0.05\; J$ could be excited
by means of optimizing the pulse shape and using a small concentration of atoms, still, 
around 100 eigenstates can be found in this tiny interval 
 even in a relatively small $40\times 40\times 40$ lattice, containing 64,000 lattice sites. Mixing 100 eigenstates would make 
the multifractal  properties completely invisible.   Indeed, 
 on the metallic side, states are uniformly excited over the whole sample (see Fig.~\ref{fig:energy_resolution}b), 
 and  exciting and imaging of  individual states does not seem to be possible.

Fortunately, as shown in Fig.~\ref{fig:energy_resolution}c, this problem can be avoided 
on the localized side of the transition by applying narrow laser beams of waist  $w_0<\xi$,
 and thereby reducing  the number of excited eigenstates by a factor  $\sim (\xi/L)^2$, where 
 $L$ denotes the linear size of the system. 
 These localized states are excited along the waist of the laser beam, but 
they  do not spatially overlap under the condition 
\be
\Delta \epsilon < {\rm BW} /\xi^3, 
\label {eq:ineq}
\ee 
with ${\rm BW}$ standing for the the bandwidth of spin $\uparrow$ bosons. To derive this inequality, we notice that a narrow laser beam excites 
states  only in a column of volume $ \xi^2 L$. Consequently the total number of excited states is 
$N \sim \xi^2 L \; \Delta \epsilon / \rm{BW}$, and the average spatial distance between these states is $d \sim L / N$. 
The overlap of the distinct excited states in the image is thus small if $d > \xi$, which leads us to the condition Eq.~\eqref{eq:ineq}.
This condition --- meaning that the energy resolution be better than 
the level spacing in a localization volume  ---  can be satisfied even relatively close to
the transition. 
The same condition is obtained  for the TPLSM protocol (Fig.~\ref{exp_setup}d). 
We thus  conclude that  imaging individual states is possible  on the localized side of the transition with the suggested  protocols 
as long as Eq.~\eqref{eq:ineq} remains satisfied.

\subsection*{Multifractal properties} Statistical properties of the critical state reflect the  structure of the 
localization quantum phase transition. Close to  the transition, the disorder-averaged correlation functions
$C_{\;\uparrow}^{(q)}(\r-\r') = \overline{|\psi_{\r\uparrow}|^{2q} |\psi_{{\r'\uparrow}}|^{2q} }$
display critical scaling,
\be
C_{\uparrow}^{(q)}(\r) \propto |\r|^{-y_q} Y^{(q)}(\r / \xi)\;.
\label{eq:correlscaleing}
\ee
The exponent $y_q$ here turns out to be intimately related to the Legendre transform $\tau_q$ of the multifractal spectrum $f(\alpha)$ 
\bea
{y}_q &= &2 {\tau}_q - {\tau}_{2q} + d , 
\eea
with $\alpha \equiv \frac{{\rm d}\tau_q}{{\rm d}q}$, $f(\alpha) = \alpha \,q - \tau_q$, and $d=3$\cite{MirlinReview}.

\begin{figure}
\centering
 \includegraphics[width=0.7\textwidth]{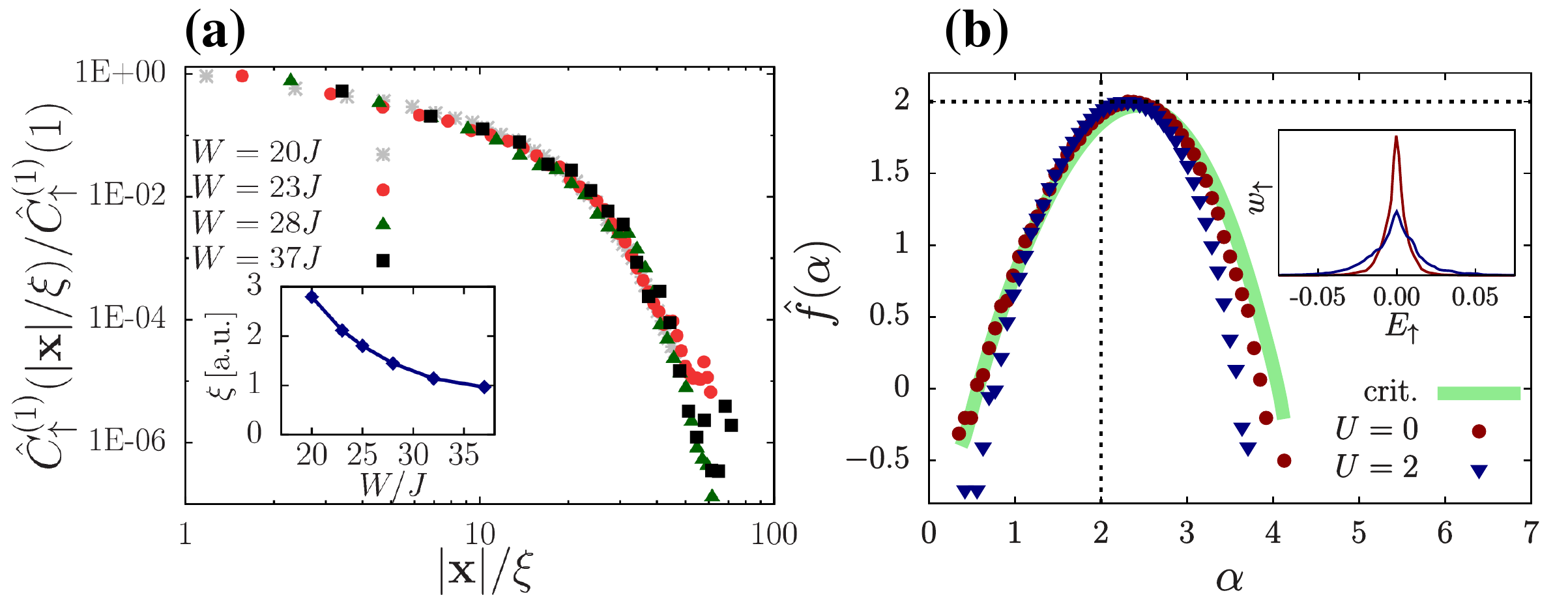}
 \caption{\textbf{(a)} Rescaled correlation function of the projected densities ${\hat\rho}_{\x\uparrow}$ for different disorder strengths 
 but fixed excitation frequency $\omega \approx 8 J / \hbar$ in the localized regime. 
 Correlations were measured along the $y$ direction in Fig.~\ref{exp_setup}.
 Other parameters were set as in Fig.~\ref{fig:energy_resolution}.
 Values of  $\xi(W)$ were determined by collapsing  of the curves.
  Inset: divergence of $\xi$  
for $W\to W_c \approx 16.5J$. \textbf{(b)} Multifractal spectrum extracted from the probability distribution of the projected density $\hat{\rho}_{\uparrow}$. 
 The measured spectrum after a pulse for non-interacting bosons (red circles) follows closely the 
 spectrum obtained from the critical state {(thick green line)}.  Interactions broaden the spectrum of the final state (see inset) and  
 slightly deform the spectrum $\hat f(\alpha)$ (blue triangles). The location of the maximum, $\alpha_{\rm max}\approx 2.3$,
characterizes the anomalous scaling of the typical density, ${\hat\rho}_{ \x\uparrow}^{\rm typ}\sim L^{-\alpha_{\rm max}}$.
 }\label{fig:MF}
\end{figure}

The previous considerations refer to  three dimensional wave functions and densities, which are quite 
difficult to determine  experimentally\cite{Nelson}.  Fortunately however, as we now show, it is also possible to observe 
the fingerprints of multifractality in   the experimentally measured \emph{projected } densities  of the gas, 
\be
{\hat\rho}_{ \x\sigma} \equiv \int dz |\psi_{\r\sigma}|^2 \; 
\ee
with $\x=(x,y)$ referring here to coordinates in the  plane  of projection.  While the rare regions of  small amplitudes (large 
$\alpha$)   appear to be washed out by the projection and multifractality is destroyed there, 
 multifractal scaling and  power law correlations are found to survive the projection
in the large amplitude (small $\alpha$) regions and are inherited by the projected wave function. For $q=1$, in particular, it is easy to prove that the projected 
densities display power law correlations with an anomalous exponent 
\be
\hat{y}_{q=1} = y_{1} - 1\approx 0.8, 
\label{eq:projection_relation}
\ee
providing a clear analytical evidence for the survival of multifractality.
This relation is indeed verified 
 by our numerics (see online Supplemental Material), which also confirms   
 that the  projected densities $\hat\rho_{\x\uparrow}$ display multifractal scaling according to Eqs.~\eqref{eq:multifractal} 
 and \eqref{eq:correlscaleing} with, however, a modified multifractal spectrum, $\hat{f}(\alpha)$, 
and modified exponents $\hat{y}_q$ and correlation functions  $\hat C^{(q)}(\x)$ for  spin $\uparrow$ bosons 
\be
\hat C^{(q)}(\x) \propto |\x|^{-\hat y_q} \hat Y^{(q)}(\x / \xi)\;.
\label{eq:correlscaleing2D}
\ee
The universal collapse  \eqref{eq:correlscaleing2D} of the correlation functions, as extracted from the horizontally projected image 
after the excitation pulse is displayed in Fig.~\ref{fig:MF}a.  As stated earlier,  one can image individual eigenstates
on the localized side of the transition  and verify  the critical scaling  of projected density-density correlations
close to the localization transition. While it is rather hard to extract the exponent 
$\hat y_{q=1}$ precisely from $\hat C^{(q)}(\x)$, the collapse onto a universal crossover 
function and the divergence of the  localization length upon approaching $W_c$ are both clear.

Density-density correlations are found to be  sensitive to the system size. In contrast, 
the multifractal spectrum characterizing the probability distribution of the projected wave functions  seems to be much more robust. 
We extracted this latter simply by measuring the distribution 
of $\ln(\rho_{\x\uparrow})$ after a pulse targeting the critical state.
The final results are shown in Fig.~\ref{fig:MF}b. 
In the absence of interactions, the spectrum of the state reached after the Rabi pulse 
follows very closely the spectrum of the true critical state itself. The maximum of  $\hat f(\alpha)$  characterizes  the  
scaling of the typical wave function  amplitudes at criticality, and the fact that $\alpha_{\rm max}>2$ is a clear evidence of the fractal nature 
of the projected wave function. 
Notice that for $\alpha\gtrsim \alpha_{\rm max}^{3D}-1\approx 3$ we do not 
expect a true multifractal spectrum: 
 large values of $\alpha$ correspond to tiny wave function amplitudes which would be mixed with larger, typical wave function amplitudes  upon
 projection, leading to the destruction of multifractality in the large $\alpha$ region.  
 Indeed,  a finite size analysis seems to confirm that the $\hat f(\alpha)$ 
function is well defined only below a critical value of $\alpha$. 

Having the spectra $\hat f(\alpha)$ (and $f(\alpha)$) at  hand, we can also perform a Legendre transformation to
determine the critical exponents $\hat \tau_q$ and $\hat y_q$ ($\tau_q$ and $y_q$), and compare them 
to the exponents extracted directly from the correlation functions (see online Supplemental Material). 
In particular, 
for $q=1$ we obtain $y_1\approx 1.8$ and $\hat y_1 \approx 0.8$, in good agreement with the exponents extracted from the scaling 
collapse of the density correlation functions, thereby verifying   the exact relation 
\eqref{eq:projection_relation}.

We need to mention two experimental aspects that may reduce the visibility of the multifractality
of the excited condensate. High resolution imaging techniques usually perform strong measurements on the position of individual 
atoms.\cite{Nelson,GreinerMicroscope,ZwierleinMicroscope}
One typically cannot therefore access $\hat{\rho}_{\x\uparrow}$ through a single shot measurement, rather, one 
has to prepare the same wave function multiple times and average the measured densities to obtain  $\hat{\rho}_{\x\uparrow}$. 
High density regions, in particular, can be measured accurately, while low density regions are less visible. 
Fortunately, the  multifractal spectrum depends only \emph{logarithmically} on these statistical fluctuations. Therefore, 
as shown in the Supplemental Material online,
already  hundreds of measurements on the same wave function can  be sufficient to extract  $\hat{\rho}_{\x\uparrow}$ and reveal  
$\hat f (\alpha)$. Particle number fluctuations may also lead to trouble: such fluctuations influence the background potential 
and lead to a slight detuning of the final state in each measurement. As  shown in the Supplemental Material online, below few percent particle number 
fluctuations are, however, not detrimental, and do not destroy the multifractal spectrum. We remark that post-selection is a natural possibility to reduce 
the error induced by particle number fluctuations.

\begin{figure}
\centering
\includegraphics[width=0.5\textwidth]{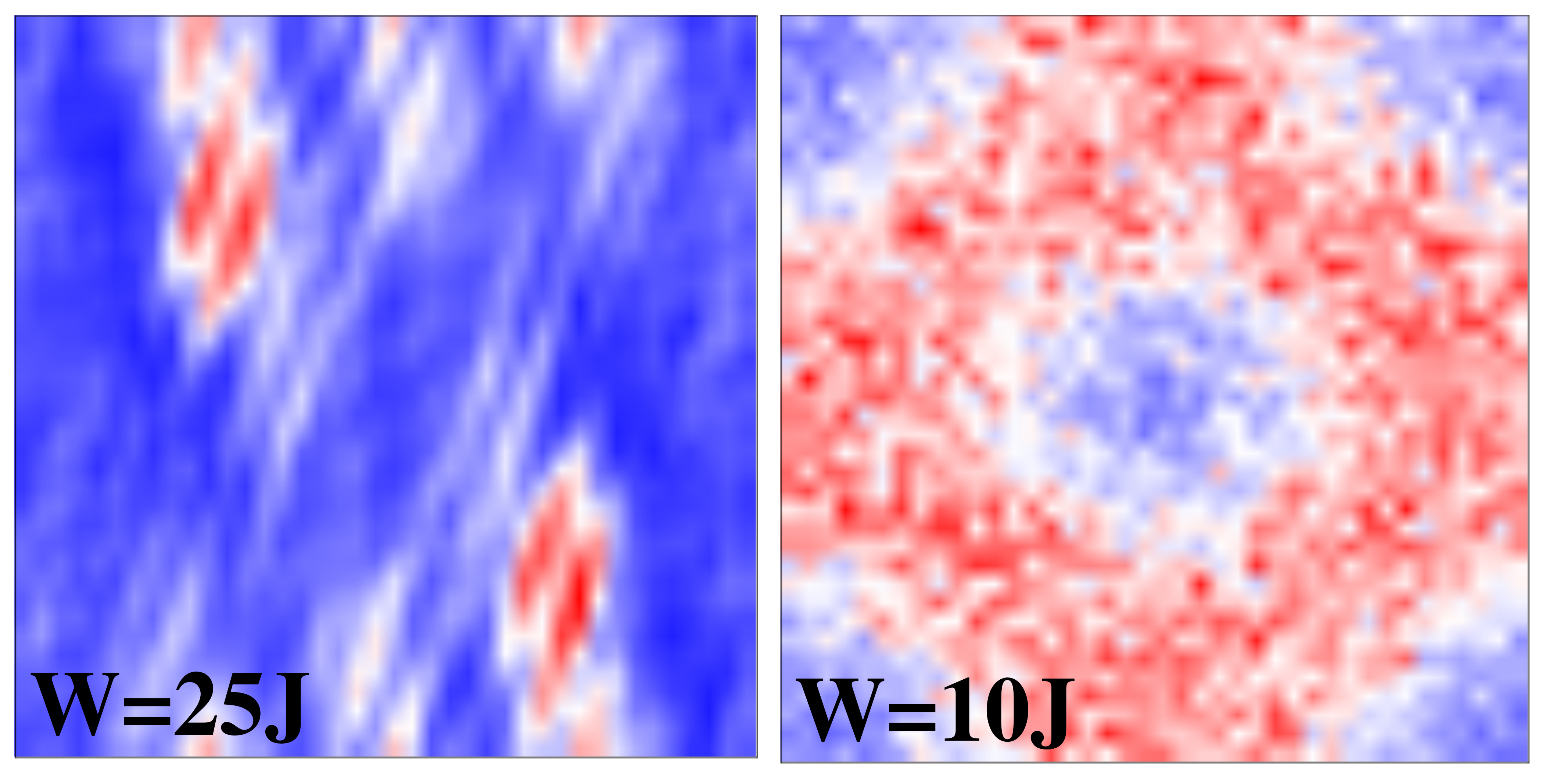}
 \caption{\label{fig:TOF}
 Time of flight image of excited atoms in the localized  (left) and in the delocalized phase 
 (right). In the localized phase interference fringes emerge, while in the delocalized phase a "Fermi surface" structure tracing smeared equal energy surfaces in momentum space appears. }
\end{figure}

\subsection*{Time of flight images}
While time of flight images do not show any particular features at criticality, they do exhibit peculiar structures 
on both sides of the localization transition. For strongly localized final states, in particular, excited atoms happen to 
localize at only a few resonant localized states. Releasing the atoms from the trap then leads to interference fringes
created by atoms released from these localized states,   similar to the
fringes observed in the case of  split condensates\cite{Schmiedmayer} (see Fig.~\ref{fig:TOF}). 
In the delocalized phase, on the other hand, weak disorder leads to the mixing of states on a relatively narrow energy shell, 
and, correspondingly,  "Fermi surface"-like  features emerge in the time of flight image, tracing 
approximately the equienergetic surfaces in the former Brillouin zone. 
 
\section*{Discussion}
Here we proposed an experimental protocol that makes it possible to image and observe 
  the so far elusive Anderson critical state using weakly interacting ultracold  
 bosons with two hyperfine states. 
  To reveal the critical correlations and amplitude fluctuations, one needs to excite 
 atoms into a few eigenstates in an extremely narrow energy window, otherwise the fragile  multifractal structure becomes completely 
 invisible.  We performed detailed simulations to verify  the  feasibility of the proposed experiment. 
Three important ingredients have been employed  to reach  the  energy 
resolution needed: (1) interactions  in the final hyperfine state were suppressed, (2)
  the shape of the excitation pulse has been optimized, and (3) excitation beams narrower than the localization length have been combined with 
  a horizontal imaging  to reduce the  number of excited eigenstates  on the localized side of the transition. 
 We demonstrated that combining these three ingredients, one can reach  energy resolutions sufficient   to  observe 
 even single localized states, and to detect   critical multifractal correlations as well as  multifractal  amplitude fluctuations, 
  even after projecting the image of the excited cloud.
 
 We have verified that the projected image inherits the multifractal structure. In particular, we have determined 
the   multifractal spectrum  of the image in the case of orthogonal symmetry, and predicted and have demonstrated by our simulations 
a universal power low  scaling  for the projected density-density correlations as a clear signature of multifractality.  Time of flight images of the excited states 
have also been shown to display  striking features in both  phases:  strong 
 localization-induced interference fringes appear in the localized phase, while momentum space equal energy surfaces can be imaged on 
 the delocalized side of the transition.
 
 Cold atomic systems  and the method proposed here thus provide a  unique framework to  reach the  critical state of 
Anderson localization and to study its multifractal properties, observed solely numerically 
 so far.  The proposed method experiment is, of course, not restricted to Anderson's localization transition. It can also  
 be used to shed light on  the critical state of
 disordered quantum Hall systems\cite{BlochEffMagnField}, disordered systems with spin-orbit coupling\cite{Spielman,MildenbergerEvers2007}, or even more exotic  topological 
 phase transitions\cite{Nomura, Moore}. The  crossed beam two-photon laser scanning imaging method would allow one
  to create and study local excitations. In particular, in the case investigated here it allows to  
  excite  single localized eigenstates and to study their spatial structure in detail. 

\section*{Methods}
\subsection*{Gross-Pitaevski simulation} 
The time evolution of the collective wave function is described by the mean field Gross-Pitaevskii equation,
\be \label{eq:GPequation}
i \partial_t \psi_{\r \sigma} = \frac{\partial H\left( \lbrace \psi_{\r \sigma}^*, \psi_{\r\sigma} \rbrace \right)}{\partial \psi_{\r\sigma}^{*}}\; .
\ee
Substituting the Hamiltonian (Eq. \eqref{eq:Hamiltonian}) in the above expression we arrive at the equations
\bea
i \partial_t \psi_{\r, \uparrow} &=& - J \sideset{}{'} \sum_{\r'} \psi_{\r',\uparrow} + \left( V_\uparrow(\r) - \omega/2 \right) \psi_{\r,\uparrow} +
\Omega_{\r}(t)  \psi_{\r,\downarrow} +  U_{\uparrow \downarrow} \left|\psi_{\r,\downarrow} \right|^2  \psi_{\r,\uparrow} 
\nonumber \\ &\equiv& \sum_{\r'} \hat  H_{\uparrow}(\psi)_{\r\r'}\psi_{\r',\uparrow} + \Omega_{\r}(t)  \psi_{\r,\downarrow}
\; ,\nonumber \\
i \partial_t \psi_{\r, \downarrow} &=& - J \sideset{}{'} \sum_{\r'} \psi_{\r',\downarrow} +  \frac{\omega}{2}  \psi_{\r,\downarrow} + 
\Omega_{\r}(t)  \psi_{\r,\uparrow} + 
 \left(U_{\uparrow \downarrow} \left|\psi_{\r,\uparrow} \right|^2 + U_{\downarrow \downarrow} \left|\psi_{\r,\downarrow} \right|^2\right)   \psi_{\r,\downarrow} 
\nonumber \\ &\equiv& \sum_{\r'} \hat  H_{\downarrow}(\psi)_{\r\r'}\psi_{\r',\downarrow} + \Omega_{\r}(t)  \psi_{\r,\uparrow}
 \; .
\eea
with the primes indicating summations over neighboring lattice sites.
In our simulations, the time-derivative of the wave function is dominated by a rapidly changing overall phase that can be easily removed by the gauge
transformation 
\be \label{eq:GPgauge}
\Phi_{\r \sigma} (t) \equiv e^{i \int_0^t E_{sc} (\Psi) dt'} \psi_{\r \sigma} (t) \; ,
\hspace{1cm}
\text{with} \hspace{1cm}
 E_{sc} (\Psi) = \frac{ \sum_{\sigma,\r} \psi_{\r \sigma}^* i \partial_t \psi_{\r \sigma}}
{\sum_{\sigma,\r} \psi_{\r \sigma}^*  \psi_{\r \sigma}} \; .
\ee
The equation of motion for $\Phi_{\r \sigma} (t)$ is then solved numerically by using a standard  4th-order Runge-Kutta scheme.

\subsection*{Chebyshev method for spectral properties}
To get spectral information about the final state $\Phi_{\r \uparrow} (t_{\mathrm{end}})$ the kernel polynomial method is used\cite{WeisseKPM}.
At the end of the excitation process the external RF field is turned off, $\Omega_{\r}(t_{\mathrm{end}}) = 0$, and the mean-field Hamiltonian has the diagonal form
\be
\hat{H}(\psi) = \left( \begin{array}{cc} \hat{H}_\uparrow(\psi) & 0 \\ 0 & \hat{H}_\downarrow(\psi) \end{array} \right) \; .
\ee
The single particle density of states for the $\uparrow$ bosons is approximated by the series
\be \label{eq:ChebyDOS}
\rho_{\uparrow}(E_\uparrow) = \sum_{\alpha}  \delta(E_\uparrow - E_\alpha)
 \approx \sum_{n=0}^{n_{\max}} g_n^{(n_{\max})} \mu_n \; T_n\left(2\;\frac{E_\uparrow-{\rm BC}}{\rm BW} \right) \; ,
\ee
where $E_\alpha$  denotes the energy of  the eigenstate  $\ket{\alpha}$  of $\hat{H}_\uparrow(\psi)$,
$T_n(x)$ stands for  the $n$'th Chebyshev polynomial and $\mu_n$ denote the Chebyshev expansion coefficients, 
which can be efficiently determined  using the recursion relation for the Chebyshev polynomials\cite{WeisseKPM}.
The spectrum is transformed to the $[-1,1]$ interval by shifting energies to the band center, 
$\rm BC$ and normalizing them by the bandwidth $\rm BW$  of the mean-field Hamiltonian. The factors 
$g_n^{(n_{\max})}$ are the expansion coefficients of the so called kernel function\cite{WeisseKPM} that regularizes the finite order approximation of the $\delta$-functions.  
In our calculations we used the Lorentz kernel $g_n^{(n_{\max})} = \sinh(\lambda (1 - n/n_{\max})) / \sinh(\lambda)$  that 
guarantees a positive definite density of states\cite{WeisseKPM}. The parameter of the kernel was set to $\lambda = 0.1$. 

A similar calculation is performed to approximate the spectral distribution of the final state of the Gross-Pitaevskii simulation
yielding
\be  \label{eq:ChebyW}
w_{\uparrow}(E_\uparrow) = \sum_{\alpha}| \scalarprod{\alpha}{\Phi_\uparrow}|^2 \delta(E_\uparrow - E_\alpha) \approx \sum_{n=0}^{n_{\max}} g_n^{(n_{\max})} \nu_n  T_n\left(2\rm \frac{E_\uparrow-{\rm BC}}{\rm BW} \right) \; ,
\ee
with  $\ket{\Phi_\uparrow}$ denoting the wave function of the $\uparrow$ bosons in the final state. The cutoff was set  to 
values as large as $n_{\max} = 30,000$  to resolve the sharp spectral peak of the final state. 

\subsection*{Data availability}
The data sets generated during and/or analysed during the current study are available from the corresponding author upon reasonable request.

\section*{Acknowledgements}
We would like to thank L. Ujfalusi and I. Varga for important and fruitful discussions.
This  research  has  been  supported  by  the  Hungarian Scientific  Research  Fund OTKA  under  Nos.  K105149.

\section*{Author contributions statement}
M.A.W. carried out the analytical and numerical calculations; G.Z., E.D., and A.A. conceived the project; 
G.Z. carried out part of the calculations and coordinated the project; M.A.W., G.Z., A.A., and E.D. prepared the manuscript.
\section*{Additional information}

\textbf{Competing financial interests:} The authors declare no competing financial interests. 

\clearpage

\clearpage
\setcounter{equation}{0}
\setcounter{figure}{0}
\setcounter{table}{0}
\setcounter{page}{1}

\makeatletter
\captionsetup[figure]{labelfont={bf},labelformat={default},labelsep=period,name={Supplementary Figure}}
\renewcommand{\thefigure}{S\arabic{figure}}
\renewcommand{\theequation}{S.\arabic{equation}}
\begin{center}
\Large{\textbf{Supplemental Material to ``Selective final state spectroscopy and multifractality in two-component ultracold 
Bose-Einstein condenstates: a numerical study''}}
\end{center}
\vspace{32pt}

\section*{Numerical analysis of projected critical eigenstates}

As a reference, we first compared the  properties of the wave functions obtained by our time dependent Gross-Pitaevskii simulation  
with those  of the critical  eigenstate of the 3D Anderson model,
\be
\hat{H}_{\mathrm{AM}} = - J \sideset{}{'} \sum_{\r, \r'} a^\dag_{\r} a_{\r'} + \sum_{\r} \varepsilon_{\r} a^\dag_{\r} a_{\r} \; .
\label{eq:anderson}
\ee
Here the prime indicates that the first sum runs over nearest neighbor sites on a 3D cubic lattice only. The operators $a^\dag_{\r}$ 
($a_{\r}$) create (annihilate) a particles at site $\r$, and  the on-site energies $\varepsilon_{\r}$ are uniformly distributed in 
the interval $ \left[ - {W}/{2}, {W}/{2} \right]$. 
We used  the  JADAMILU library\cite{ref_jadamilu} to extract the exact critical eigenstates of the Anderson model 
up to linear system sizes $L = 120$ (i.e. $\sim 1.7\; 10^{6}$ sites), where length is measured in units of the lattice constant $a$, and analyzed their multifractal properties.

\begin{figure}
\centering
 \includegraphics[width=\textwidth]{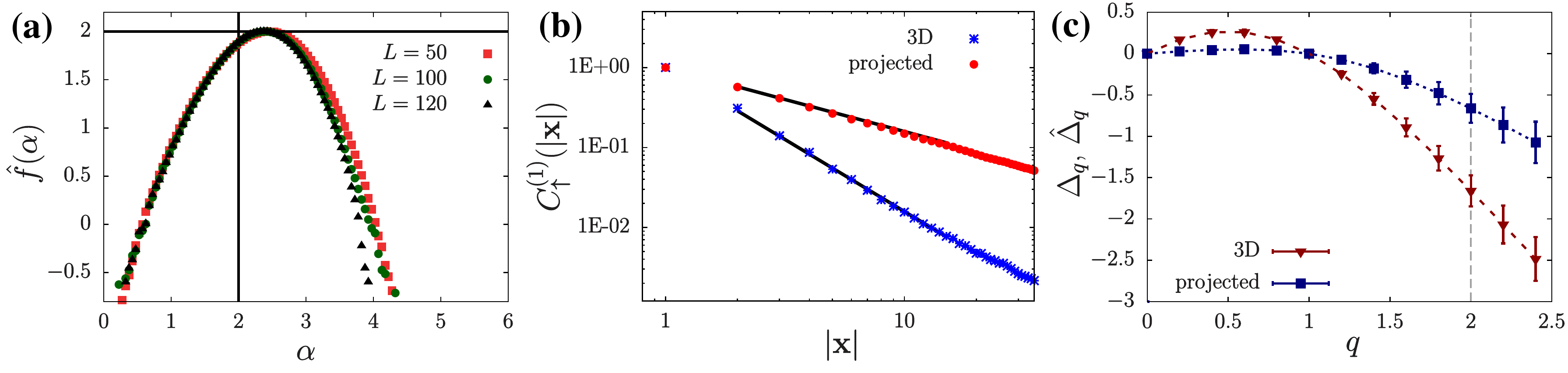}
 \caption{ Multifractal spectrum of the projected critical states of an ensemble of disordered samples with  $W_c/J = 16.5$, and linear system sizes up to $L=120$.
  \textbf{ (a)} Multifractal spectra $\hat{f}(\alpha)$ calculated from the projected density for different system sizes.
 The high-amplitude (small $\alpha$) part of the curves are found to be system-size independent. At the low-amplitude (large $\alpha$) part 
 finite size effects make the spectra slightly system-size dependent. \textbf{(b)} Density-density correlations  for the 
 three-dimensional density and the projected density of the critical state. Contiunous black lines show power-law fits with exponents 
 $y_1 = 1.82(1)$ and $\hat{y}_1 = 0.832(3)$. \textbf{(c)} Anomalous dimensions $\Delta_{q}$ and $\hat{\Delta}_{q}$ for the 
 three-dimensional  and the projected densities,  respectively, extracted from the Legendre transforms of the 
 corresponding $f(\alpha)$ and $\hat f(\alpha)$ functions. The relations $y_1 = -\Delta_{2}$ and $\hat{y}_1 = -\hat{\Delta}_2$ 
  are satisfied within error bars.}
 \label{suppfig_crit}
\end{figure}

Having obtained the exact critical eigenstates of \eqref{eq:anderson}, we   determined  their  projected densities $\hat{\rho}_{\x}$. 
Assuming that  $\hat{\rho}_{\x}$  displays $\hat d=2$ dimensional multifractality,   
we expect its probability density to scale as 
\be
P(\hat{\rho}_\x \sim L^{-\alpha}) \propto L^{\hat{f}(\alpha) - \hat d} \; ,
\label{eq:projmulti}
\ee
with  $\hat{f}(\alpha)$ denoting  the multifractal spectrum
of the projected density. 
To test the validity of Eq.~\eqref{eq:projmulti} and the size independence of the multifractal
spectrum $\hat f(\alpha)$,  we computed  an ensemble of critical eigenstates and extracted  $\hat{f}(\alpha)$  from it 
for system sizes in the range of $50\le L \le  120$. 
The results summarized in Fig. \ref{suppfig_crit} evidence  the system-size independence of $\hat{f}(\alpha)$
for $\alpha\le 2.4$, while   the maximum being shifted from $\alpha = 2$ indicates the 
anomalous scaling of the typical projected density,  $\hat{\rho}_{\mathrm{typ}} \sim  L^{-\alpha_{\rm max}}$.

As shown in Fig. \ref{suppfig_crit},
both the $d=3$  dimensional and the projected $\hat d = 2$- dimensional critical wave function  densities 
display  power-law correlations
 \be
C^{(1)}(\r) \propto |\r|^{-{y_1}} \; , \quad \mathrm{and} \quad \hat{C}^{(1)}(\x) \propto |\x|^{-{\hat{y}_1}} \; ,
\ee
with the  critical exponents of the correlation functions are related as 
\be
\hat{y}_1 = y_1 - 1 \; .
\label{eq:y1_hat_y1}
\ee
To verify this connection we rewrite the correlation function of the projected densities as
\be
\hat{C}^{(1)}(\x) = \overline{\hat{\rho}_{\x} \hat{\rho}_{\mathbf{0}}} = \int dz \int dz' \;  \overline{\rho_{\r} \rho_{\r'}} =
\int dz \int dz' \;  C^{(1)}(\r-\r') \; ,
\ee
where $\r = (x,y,z)$, $\r' = (0,0,z')$, and disorder averaging is denoted by  overline. Inserting the critical three-dimensional correlation function 
yields 
\be
\hat{C}^{(1)}(\x) = \int dz \int dz' \; \bigl(|\x|^2 + (z-z')^2\bigr)^{-y_1/2} \sim |\x|^{-y_1 + 1} \int d Z \bigl(1 + Z^2\bigr)^{\; -y_1/2} \sim |\x|^{-y_1 +1} \; ,  
\ee
where in the last step we assumed  $|\x|\ll L$ and used $y_1\approx 1.8 > 1$ to carry out the integral. 
As shown in Supp. Fig. \ref{suppfig_crit}, our numerics are compatible with the simple analytical result, \eqref{eq:y1_hat_y1}.

As  mentioned in the main text, the correlation exponents $y_q$  and $\hat{y}_q$ can also be determined from the  
multifractal spectra $f(\alpha)$  and $\hat{f}(\alpha)$  (see Eqs. (5) and (6) in the main text and  Ref.~\cite{ref_MirlinReview}).
Taking the Legendre-transform of $f(\alpha)$ [ $\hat{f}(\alpha)$] yields $\tau_q = 3(q-1) + \Delta_{q}  $ 
[$\hat \tau_q = 2(q-1) + \hat\Delta_{q}  $]. 
Panel (c) in Supp. Fig. \ref{suppfig_crit} shows the resulting anomalous dimensions $\Delta_{q}$ and $\hat{\Delta}_{q} $ for the three-dimensional 
and the projected two-dimensional densities,  respectively. 
According to Eq. (6) in the main text, the relations to the exponents $y_q=2\tau_q -\tau_{2q}+d$  and $\hat y_q=2\hat \tau_q -\hat \tau_{2q}+\hat d$
imply  $y_1 = -\Delta_2$ and $\hat{y}_1 = -\hat{\Delta}_2$, respectively.
Our numerical results in panel (c) of Supp. Fig. \ref{suppfig_crit} indeed  support these relations.

\section*{Three dimensional multifractality of the Gross-Pitaevskii wa\-ve function}
\begin{figure}
\centering
 \includegraphics[width=0.7\textwidth]{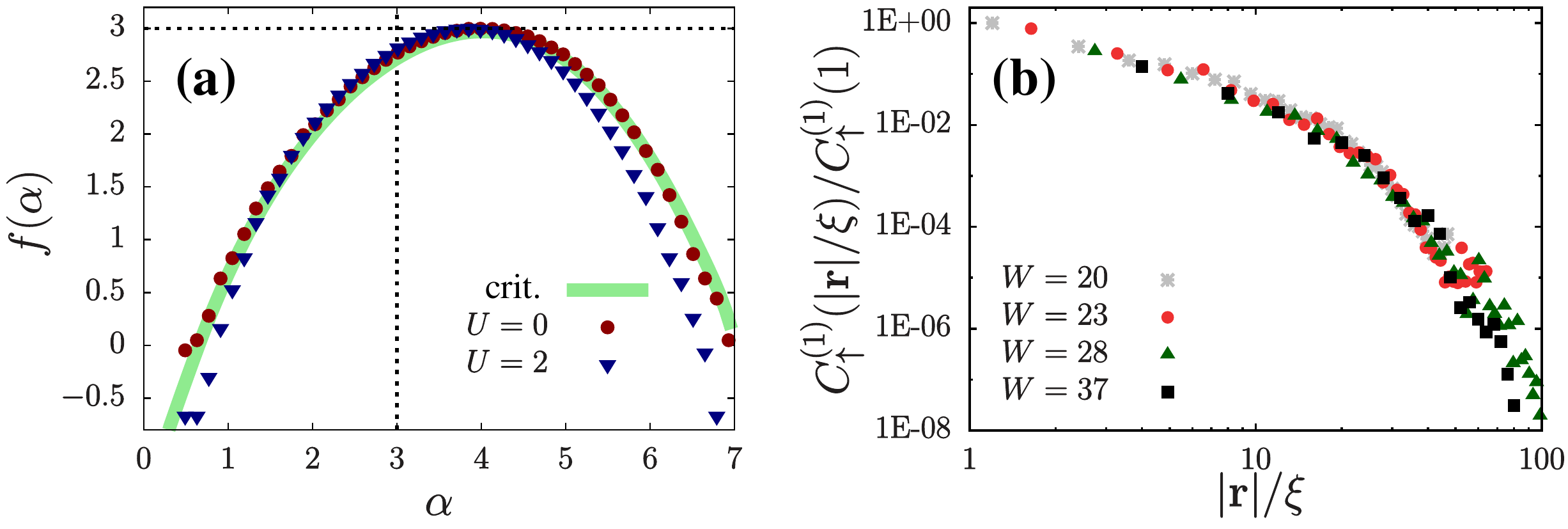}
 \caption{\textbf{(a)}  Multifractal spectrum extracted from the probability distribution of the three-dimensional density $\rho_{\r\uparrow}$. 
 The spectrum measured  after a pulse for non-interacing bosons (red circles) follows closely the 
 spectrum obtained from the critical state {(thick green line)}.  
 Interactions  mix in several eigenstates and 
 deform the spectrum $f(\alpha)$ in the region of rare events (blue triangles).  
 \textbf{(b)}  Rescaled correlation function of the three-dimensional densities 
 $\rho_{\r \uparrow}$ for different disorder strengths 
 but fixed excitation frequency {$\omega \approx 8 J / \hbar$} in the localized regime. }\label{suppfig_3D}
\end{figure}

Multifractal properties of the $d=3$ dimensional 
 Gross-Pitaevskii wave function have also been tested with findings 
 similar to the ones presented in the main text~\cite{ref_paper}. 
 The spectra in  Supp. Fig. \ref{suppfig_3D}
 were  determined  from the distributions of the Gross-Pitaevskii densities   $\rho_{\r,\uparrow}$.  In the three-dimensional case box-averaging is necessary to reveal the 
 universal multifractal spectrum\cite{Romer}. For the Gross-Pitaevskii wavefunction we used $L=40$ and box sizes $l=2$, 
 while for the critical eigenstates $L=100$  and $l=5$, yielding  the same  effective system size  $L/l=20$ in both cases. 
 Similar to the projected densities' multifractal spectrum, the high-amplitude (small $\alpha$) part of the 
  Gross-Pitaevskii  multifractal  spectrum is very close to the one extracted from single critical eigenstates,
  in spite of the presence of interactions during time evolution. 
Deviations in the low-amplitude (large $\alpha$) part may be explained by the fact that the Gross-Pitaevskii wave function is 
not a single eigenstate, but  a superposition of several eigenstates of the self-consistent Hamiltonian.
The $d=3$ dimensional density-density correlations of the cloud 
can also be collapsed  on the localized side of the transition by just rescaling the length and the wave function amplitudes
(see  Supp. Fig. \ref{suppfig_3D}~(b)).

 \section*{Strong measurement of the projected atom density}
 
 As briefly discussed in the main text\cite{ref_paper}, direct measurement of the projected density $\hat{\rho}_{\x\uparrow}$  is difficult because 
 most imaging techniques perform strong quantum measurements on the spatial positions of the individual atoms.
 Therefore atoms appear as dots in the images. 
 In a given measurement the expectation value  of the atoms that are found in the projected position $\x$ is simply $\hat{\rho}_{\x\uparrow}$. 
 Since  -- assuming Poissonian statistics --  the relative fluctuation of the measured particle number at site $\x$ from its average  is simply $\hat{\rho}_{\x\uparrow}^{-1/2}$,   measuring small densities accurately in single-shot experiments is impossible. One can overcome this difficulty by preparing the same condensate wave function multiple times. 
 The expectation  value of the total number of atoms at the projected position $\x$ is then $N_M \hat{\rho}_{\x\uparrow}$, with $N_M$ denoting the total number of measurements. Thereby,   relative fluctuations are reduced to $\left( N_M \hat{\rho}_{\x\uparrow} \right)^{-1/2}$.
Fortunately, since  $\alpha$ is a logarithmic function of the density, the spectrum $\hat f(\alpha)$ can be reconstructed even 
from a rough estimate of $\hat{\rho}_{\x\uparrow}$ 
 with only 10-20\% relative error.

\begin{figure}
\centering
 \includegraphics[width=0.8\textwidth]{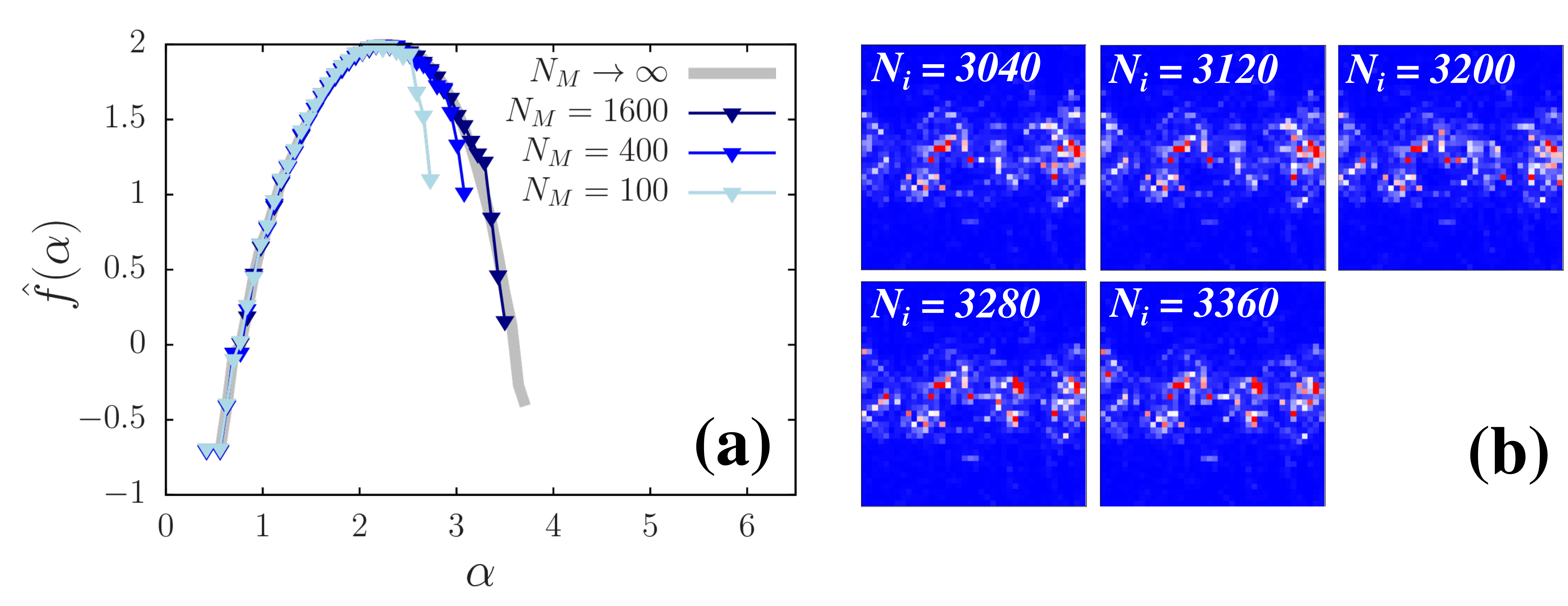}
 \caption{{\textbf{(a)} Reconstruction of the $\hat f (\alpha)$ multifractal spectrum from different number ($N_M$) of emulated strong measurements. 
  While already $N_M \approx 100$ measurements  reveal the small $\alpha$ part of the function, with $N_M \gtrsim 1000$ measurements almost the 
  whole $\hat f (\alpha)$ becomes visible.
 \textbf{(b)} Projection images from the final condensate with different initial particle numbers $N_i$, while keeping the disorder potential, and RF frequency fixed. 
 The final state and thus the multifractal spectrum remain almost unchanged even for particle number fluctuations of  10\%.}}\label{suppfig_strong}
\end{figure}

 We emulated the statistical  effects of strong quantum measurements on the Gross-Pitaevskii wave function by assuming  an independent Poisson 
 distribution  of expectation value  $\langle N_\x\rangle =\hat{\rho}_{\x\uparrow}$  for the measured number of atoms, $N_\x$ at each  projected position $\x$.
  The measured single shot densities were then averaged, and the 
 $\hat f(\alpha)$ multifractal spectrum was calculated from the averaged density. As  shown in panel (a)  of Fig. \ref{suppfig_strong}, performing 
 roughly 1,000 single-shot measurements for identically prepared condensates is enough to reveal the desired $\hat f(\alpha)$ function for $\alpha \lesssim 3$, i.e.  to see clearly the nontrivial  
 position $\alpha_{\max}$ of the maximum of $\hat f(\alpha)$. However, already $N_M=100$ single shot experiments are  enough to measure the entire  projected spectral function in the regime $\alpha\lesssim 2,4$. 
 
 To perform multiple measurements on the same state, one must be able to prepare the same condensate wave function multiple times. The nonlinearity induced by the $U_{\uparrow \downarrow}$ and $U_{\downarrow \downarrow}$ interaction terms makes the quantum state  sensitive to the 
 total number of atoms. Since fluctuations in the total atom number are unavoidable, we also tested the stability of the final multifractal spectrum 
 against particle number fluctuations. As demonstrated in panel (b) of Fig. \ref{suppfig_strong},  even a 10\% change  in the initial density leaves the final state almost untouched.

\end{document}